\documentclass[aps,pre,reprint,amsmath,amssymb,nofootinbib]{revtex4-2}

\usepackage{graphicx}
\usepackage{bm}
\usepackage[colorlinks=true,citecolor=blue,linkcolor=blue,urlcolor=blue]{hyperref}
\makeatletter\let\auto@bib@innerbib\@empty\makeatother

\begin{document}

\title{Separating Field--Camp Alignment from Structural Balance in a Frustrated Random-Field Ising Model of Social Networks}

\author{Natalia Romero}
\affiliation{Machine Perception and Cognitive Robotics (MPCR) Laboratory, Florida Atlantic University\\Natural and Artificial Intelligence Institute (NAII)}

\begin{abstract}
Frustrated Ising models of social networks mix ferromagnetic ($J_{ij}>0$, agreement-seeking) and
antiferromagnetic ($J_{ij}<0$, antagonistic) couplings, and the antagonistic bonds are almost always
placed by drawing each bond's sign independently at a fixed rate --- which yields a $\pm J$ spin
glass. Real signed networks (Epinions, Slashdot, Wikipedia) instead show a consistent excess of
balanced triads over what independent placement produces, and a perfectly balanced placement is
gauge-equivalent to an unfrustrated ferromagnet. We compare three sign-placement conditions on a
frustrated random-field Ising model of a polarised population, sharing an identical graph, field assignment, and edge-weight magnitude: an independent
baseline; a balance-structured condition whose camp membership coincides with each voter's own field
sign; and the same balance-structured condition with camp labels randomly permuted across voters, so
that triadic balance is identical to the aligned condition by construction while the camp--field
correlation is removed. On a confound-controlled Kolmogorov--Smirnov test of the sampled energy-gap
distribution, sampled by classical simulated annealing, the aligned condition differs from the
independent baseline at every density tested
($p=7.3\times10^{-8}$), while the permuted condition --- despite identical triadic structure --- shows
no significant difference from the baseline (paired mean difference $0.003$, 95\% CI
$[-0.041,0.046]$) and remains sharply different from the aligned condition. These results isolate
field--camp alignment, not triadic balance considered alone, as the quantity responsible for the
effect. Because exact balance sits at the unfrustrated, ferromagnet-equivalent end of this range, this
qualitative direction is an expected consequence of a known mapping; the contribution here is the quantitative,
confound-controlled measurement of the effect's size away from that limit, shown to survive
parallel-tempering sampling, hold at system sizes up to $N=300$, and persist under a structural-noise
level calibrated to real-network statistics, while triadic balance alone produces no detectable
effect at $N=90$ and at most a small one at $N=180$--$300$ (paired $\Delta D_{KS}\approx0.04$, 95\%
CI upper bound $<0.09$), well below the alignment effect.
\end{abstract}

\maketitle

\section{Introduction}

Spin glass models were introduced to describe disordered magnetic materials with competing
interactions \cite{binder1986,mezard1987}; their defining feature is \emph{frustration} --- bonds
favouring alignment and bonds favouring anti-alignment coexisting on a graph that cannot
simultaneously satisfy both. Applied to social systems, antagonistic (``contrarian'') bonds are the
standard device for introducing frustration into an otherwise consensus-seeking population
\cite{galam2004}, and the broader mapping between opinion dynamics and statistical physics is well
established \cite{castellano2009}. Almost universally in this literature, the sign of each bond is
drawn independently at a fixed rate, with no structural dependence on the rest of the network.

Real signed social networks --- ``signed'' in the network-science sense \cite{harary1953}, each tie
labelled positive or negative, the counterpart of a mixed-sign coupling $J_{ij}=\pm|J_{ij}|$ --- are
not built this way. Across Epinions, Slashdot, and Wikipedia
request-for-adminship data, the empirical frequency of balanced triads (all-positive, or
two-negative-one-positive) substantially exceeds what independent sign placement at the same
marginal rate would produce \cite{leskovec2010} --- a pattern structural balance theory predicts
from first principles \cite{heider1946,cartwright1956} and for which computable global measures
exist at the scale of real large networks \cite{facchetti2011}. Macy, Szymanski and Ho{\l}yst's
centennial review of the Ising model identifies signed coupling networks with an explicit empirical
parameterisation as a live open problem in the field \cite{macy2024}. A closely related research
programme has shown that triadic, homophily-based structure --- rather than independent dyadic
assignment --- is needed to reproduce real social-triad statistics \cite{korbel2023,pham2022},
motivating our own triadic treatment of sign placement below.

This raises a direct physical question: at a fixed marginal rate of antagonistic bonds, does the
\emph{structure} in which those bonds are arranged --- balanced versus independently placed ---
change the resulting energy landscape and sampled energy-gap distribution, or is marginal rate the
only quantity that matters? Sign placement sets how frustrated the model is: independent placement
gives a $\pm J$ spin glass, while perfectly balanced placement removes frustration entirely
(Sec.~\ref{sec:gauge}). The natural way to build a balanced signing from real data, however,
confounds two things that are not the same. In the calibration used throughout this paper, each
voter's camp is set by the same real partisan-lean tier that sets their field sign
(Sec.~\ref{sec:sign}), so a near-perfectly balanced signing is simultaneously a near-perfectly
\emph{field-aligned} one: triadic balance and field--camp alignment move together by construction,
and a comparison against an independent baseline cannot on its own tell them apart.
\textbf{This paper's central finding is that alignment, not triadic balance considered on its own,
is responsible for the detected effect.} Disentangling the two requires a second, further
confound-controlled design --- holding triadic structure exactly fixed (not approximately, by
construction) while permuting which voter carries which camp label, so that only the alignment
between camp and field is destroyed. Once isolated this way, triadic structure alone produces no
detectable signal at $N=90$ and at most a small one at larger $N$, while field--camp alignment
produces a large one. We establish this in two
steps: first, a confound-controlled comparison against an independent baseline (at fixed marginal
rate) shows that balanced sign placement measurably changes the sampled energy-gap distribution;
second, a further confound-controlled permutation isolates which of balance's two conflated
ingredients --- structure or alignment --- is actually responsible (Sec.~\ref{sec:isolate}).

Exact Heider balance is not itself a new physical limit: by the Cartwright--Harary partition, a
perfectly balanced signing is gauge-equivalent to an unfrustrated ferromagnet in a transformed field
\cite{mattis1976,toulouse1977,harary1953} (Sec.~\ref{sec:gauge} makes this explicit for the present
model). That same mapping, in fact, already predicts the qualitative direction of the isolation
result above: at \emph{any} level of structural noise, not only exact balance, the aligned and
permuted conditions reduce to the same transformed bond disorder under a uniform field versus a
random-sign field respectively --- a known qualitative picture (Sec.~\ref{sec:gauge}) --- so that
alignment ordering more than permutation is an expected consequence of a known mapping, not a new
qualitative discovery. The question this paper answers is therefore neither whether the balanced
endpoint is ordered (it must be) nor even whether alignment beats structure qualitatively (the
mapping already implies it), but how large that effect is in the specific statistic this paper
measures, whether that size is statistically detectable above this finite system's sampling noise,
and whether it survives equilibrium sampling, calibration to real-network statistics, and
multiple-comparisons correction.

\textbf{Contributions.} (1) \textbf{We isolate field--camp alignment from triadic structure
directly} (Sec.~\ref{sec:isolate}) --- randomly permuting camp labels across voters while reusing the
same realised edge-sign noise, which leaves every cycle's sign product, and therefore triadic
balance, exactly unchanged by construction (not an approximation confirmed only by measurement)
while reducing the correlation between camp and each voter's own field sign to approximately zero
--- and show no significant difference from the negative control is detected in the permuted
condition's sampled energy-gap distribution, while a direct check of the lowest observed
configuration shows substantially lower bond satisfaction once alignment is removed, confirming this
is not a cosmetic relabelling even though the bond network's intrinsic frustration is unchanged.
Triadic structure held fixed, on its own, does not reproduce a detectable signal at this sample size;
field--camp alignment does. This result is reproduced using parallel-tempering sampling
(Sec.~\ref{sec:pt-extension}), and holds when the balance-structured condition's own structural-noise
level is calibrated to measured real-network balance statistics rather than an arbitrary endpoint
(Sec.~\ref{sec:results}). (2) Before isolating it, we first establish the aligned effect itself: near-perfect structural
balance, at the same marginal negative-edge rate as an independent baseline, substantially changes
ground-state landscape statistics (bond satisfaction, field-dominance crossover location, sampled
magnetisation variance) at every network density tested, and we connect this directly to the known
gauge equivalence between balanced signing and an unfrustrated ferromagnet, rather than presenting
it as an unexplained empirical regularity. (3) Using a confound-controlled design and a
multiple-comparisons-corrected statistical test, we show this produces a genuinely different sampled
energy-gap distribution, not merely a mean-shift in ground-state energy --- a result that
strengthens, rather than weakens, as the model's field/coupling ratio grows, quantifying how the full
sampled population, not just the ground state, responds away from the exactly-balanced limit. (4) We
note, as a schedule-bookkeeping point rather than a separate theoretical contribution, that an
apparent weakening of the (3) signal at high field/coupling ratio under one common annealing-schedule
convention is a measurement artefact, not a real physical effect --- relevant to any study of this
model class that varies the field/coupling ratio under a fixed schedule.

\section{Model and Methods}

\subsection{Hamiltonian}

The model is $N$ voters, each a spin $s_i \in \{-1,+1\}$, governed by
\begin{equation}
H(\bm{s}) = -\sum_i h_i s_i - \sum_{(i,j)\in E} J_{ij}\, s_i s_j ,
\label{eq:hamiltonian}
\end{equation}
with the sign of $h_i$ set by individual $i$'s own partisan direction (positive $h_i$ favours
$s_i=+1$, negative $h_i$ favours $s_i=-1$, independent of neighbours; Sec.~\ref{sec:calibration} gives the per-tier
magnitudes and signs), $J_{ij} > 0$ (``ferromagnetic'') favouring aligned neighbours, and $J_{ij} <
0$ (``antiferromagnetic'') favouring anti-aligned neighbours. With both signs present the model is
a frustrated random-field Ising model: any cycle whose sign-product is negative cannot have all its
bonds satisfied at once \cite{toulouse1977}, and how much such frustration exists is set entirely by
where the negative signs are placed (Sec.~\ref{sec:sign}). Ground states and finite-temperature
samples are obtained by simulated annealing \cite{kirkpatrick1983} via \texttt{dwave-neal} 0.6.0
(\texttt{SimulatedAnnealingSampler}), a single-spin-flip Metropolis update with a geometric
inverse-temperature schedule over $\beta\in[0.1,10.0]$, 1000 reads and 1500 sweeps per instance unless
stated otherwise, each read independently initialised at a uniformly random configuration; every
reported condition in this paper uses an explicit, reproducible seed derived deterministically from a
fixed base seed, density, and realisation index (never an unseeded or wall-clock-derived draw). This
holds at every tested $r_h$, including $0.5$ and $1.5$: only the Hamiltonian's parameters change with
$r_h$, never the sampler, schedule, or seeding scheme. Throughout this paper, ``ground state'' denotes
the lowest-energy configuration observed across a run's reads, not a certified global optimum or an
equilibrium-sampled state; likewise ``energy gap'' denotes a sampled energy difference relative to
that run's own observed minimum, and its distribution across reads depends on which configurations the
sampler reaches and how often, not on the true density of states --- a sampled energy-gap
distribution, not the Hamiltonian's true excitation spectrum. The solver's native sign convention is verified
against this stated Hamiltonian by a dedicated automated test suite, including brute-force cross-checks against
exhaustive enumeration for small instances, prior to any reported result. The ratio $r_h = h/J$
(held throughout as an explicit, swept multiplier on a calibrated raw $h$ scale, $r_h=1$ being that
raw scale) governs whether the system orders collectively through its coupling structure or is set,
spin by spin, by individual field terms; we treat it as a swept axis throughout, since the central
result below (Sec.~\ref{sec:results}) depends on where the system sits on it.

\subsection{Empirical calibration of field and coupling magnitudes}
\label{sec:calibration}

Individual conviction field magnitude $|h|$ is derived from Bartels' (2000) \cite{bartels2000}
probit-coefficient analysis of party-identification strength effects on US presidential vote choice,
expressed as a percentage-point behavioural effect size relative to a 25-percentage-point ceiling
\cite{nickerson2008} via the standard probit-to-marginal-effect conversion \cite{long1997}: $|h| =
1.729$ (strong partisans), $1.114$ (weak partisans), $1.256$ (independent leaners) --- leaners
exceeding weak partisans, not a monotonic taper by identification strength, reproduces Keith et
al.'s (1992) \cite{keith1992} finding that independent leaners ``think and act rather like straight
partisans.'' Population shares by partisan-intensity tier and direction are taken from ANES 2020
party-identification data \cite{anes2020}. The \emph{sign} of $h_i$ is set by partisan direction ---
negative for Democratic-leaning tiers, positive for Republican-leaning tiers, zero for true
independents, who carry no directional lean in this calibration --- the same directional split used
for camp membership $c_i$ in Sec.~\ref{sec:sign}, so that $\mathrm{sign}(h_i)=c_i$ exactly for every
partisan voter; this correlation is directly relevant to the gauge argument of
Sec.~\ref{sec:gauge}. Coupling magnitude $|J_{ij}|$ is set by a stylised four-tier ordering by social
proximity --- household contact $0.55$, neighbour contact $0.40$, online close friend $0.25$, weak tie
$0.10$ --- sampled per edge from a fixed tier-frequency distribution. This ordering reflects the
general expectation, consistent with the social-influence literature \cite{gerber2008,bond2012,
granovetter1973}, that stronger ties couple opinions more tightly than weaker ones; it is not a
directly-measured opinion-alignment coefficient; Granovetter \cite{granovetter1973} is a
qualitative/theoretical account of tie strength with no quantitative effect-size estimate, and Gerber
et al.\ \cite{gerber2008} and Bond et al.\ \cite{bond2012} report turnout and voting-mobilisation
effects (percentage-point changes and odds ratios), not opinion-alignment effects. A transparent,
quantitatively justified calibration of $|J_{ij}|$ against opinion-specific (not turnout- or
mobilisation-specific) social-influence data remains future work. These magnitudes are held identical
across every sign-placement condition compared below; only the \emph{sign} assigned to each edge
differs.

\subsection{Network construction}
\label{sec:network}

Networks are degree-limited Barab\'asi--Albert graphs \cite{barabasi1999} with attachment parameter
$m \in \{2,5,10\}$, pruned to a maximum degree of 20, verified enforced exactly (0 violations across
densities in a 300-realisation check). The naive unpruned expectation $\langle k\rangle\approx 2m$
materially overstates the realised mean degree once pruning is active, most severely at $m=10$ (naive
$20$ vs.\ measured $12.37$, a $38\%$ overstatement); measured directly (30 realisations/density,
matching the value independently used in this model's own mean-field theory), realised mean degree is
$3.83$ ($m=2$), $8.33$ ($m=5$), and $12.37$ ($m=10$). Density is tracked as a free variable throughout.

\subsection{Sign placement: the independent baseline and the balance-structured alternative}
\label{sec:sign}

Two sign-placement rules are compared, matched at the same marginal negative-edge rate. The
\textbf{independent} rule draws each edge's sign i.i.d.\ Bernoulli($\alpha$), independent of network
structure, at rate $\alpha$ (negative) versus $1-\alpha$ (positive); all comparisons below use
$\alpha=0.50$ --- the symmetric $\pm J$ spin glass, here in a random field. The
\textbf{balance-structured} rule instead assigns each node to one of two camps
using the same real partisan-lean tiers used for $h$ (Sec.~\ref{sec:calibration}), then applies Cartwright \& Harary's
(1956) \cite{cartwright1956} two-camp structural-balance theorem directly: within-camp edges
positive, cross-camp edges negative, governed by a structural-noise parameter $\nu\in[0,1]$ ($\nu=0$
= perfect balance, $\nu=1$ = perfect anti-balance). Writing $c_i\in\{+1,-1\}$ for node $i$'s camp
label, the realised sign is
\begin{equation}
J_{ij} = \xi_{ij}\, c_i c_j\, |J_{ij}|, \qquad
\xi_{ij} = \begin{cases} +1, & \text{prob. } 1-\nu \\ -1, & \text{prob. } \nu, \end{cases}
\label{eq:balance-rule}
\end{equation}
with $|J_{ij}|$ fixed from the empirical calibration (Sec.~\ref{sec:calibration}) and $\xi_{ij}$ drawn independently
per edge. At $\nu=0$, $J_{ij}=c_ic_j|J_{ij}|$ for every edge, so every triad's sign product is $+1$
(exact Cartwright--Harary balance, and no frustration; Sec.~\ref{sec:gauge}), and raising $\nu$
reintroduces frustration edge by edge; at $\nu=1$ every balance-dictated sign is flipped (maximal
anti-balance). Because camp membership is close to $50/50$ and uncorrelated with graph structure,
$\mathbb{E}[\xi_{ij}c_ic_j]$ averages out and the balance-structured rule's marginal negative-edge
rate stays near $50\%$ across the full $\nu$ range --- matching the independent baseline's
$\alpha=0.50$ by construction, not by tuning. Matched marginal rate alone does not imply matched
triadic structure: the independent baseline's expected balanced-triad fraction is
$(1-\alpha)^3+3\alpha^2(1-\alpha)$ (binomial in the number of negative edges per triad), which
equals $0.50$ at $\alpha=0.50$, whereas the same binomial logic applied to
Eq.~\eqref{eq:balance-rule} (a triad is balanced iff an even number of its three edges are flipped
from the balance-dictated sign) gives $(1-\nu)^3+3\nu^2(1-\nu)=0.8645$ at $\nu=0.05$ --- well above
the independent baseline's $0.50$, though not itself near-total order. That gap --- same marginal
rate, sharply different triadic structure
--- is the genuine, non-tautological difference this baseline comparison holds fixed by design:
marginal rate is matched, but triadic structure and field--camp alignment both still differ between
the independent baseline and the balance-structured condition at once. Sec.~\ref{sec:isolate}
separates these two remaining, conflated ingredients directly.

\subsection{Exact balance is gauge-equivalent to a ferromagnet}
\label{sec:gauge}

By the Cartwright--Harary partition, the $\nu=0$ signing is $J_{ij}=c_ic_j|J_{ij}|$ for every edge.
The gauge transformation $\sigma_i = c_i s_i$ (so $s_i=c_i\sigma_i$, since $c_i\in\{+1,-1\}$) sends
Eq.~\eqref{eq:hamiltonian} to
\begin{equation}
H(\bm{s}) = -\sum_i (c_ih_i)\,\sigma_i - \sum_{(i,j)} |J_{ij}|\,\sigma_i\sigma_j,
\label{eq:gauge}
\end{equation}
since $J_{ij}s_is_j = (c_ic_j|J_{ij}|)(c_i\sigma_i)(c_j\sigma_j) = c_i^2c_j^2|J_{ij}|\sigma_i\sigma_j
= |J_{ij}|\sigma_i\sigma_j$, using $c_i^2=c_j^2=1$; the coupling term is an unfrustrated ferromagnet
in the transformed variables, for any graph. This is the Mattis construction
\cite{mattis1976,toulouse1977}, and the equivalence between balance and the absence of frustration
is Harary's original result \cite{harary1953}: a signed graph is balanced if and only if it is gauge
transformable to an all-positive one. Because $\mathrm{sign}(h_i)=c_i$ for every partisan voter in
this calibration (Sec.~\ref{sec:calibration}), the gauge-transformed field $c_ih_i=|h_i|>0$ is \emph{also}
uniform-sign for that subpopulation: at $\nu=0$, Eq.~\eqref{eq:gauge} is an unfrustrated ferromagnet
with every term --- field and coupling alike --- favouring $\sigma_i=+1$ for every partisan voter,
so the ground state $s_i^*=c_i$ (every voter's opinion matching their own partisan lean) is close to
built into the construction rather than a result of the dynamics. The true-independent/other
subpopulation (24\% of voters, $h_i=0$, camp label assigned uniformly at random,
Sec.~\ref{sec:sign}) introduces no ground-state disorder at $\nu=0$, despite carrying no systematic
field alignment: on a connected graph with $|J_{ij}|>0$ everywhere and at least one strictly positive
transformed field, $H(\bm\sigma)\ge H(\mathrm{all}{+}1)$ for \emph{every}
configuration $\bm\sigma$ --- each field term satisfies $-c_ih_i\sigma_i\ge-c_ih_i$ (equality iff
$\sigma_i=1$ or $c_ih_i=0$) and each bond term satisfies $-|J_{ij}|\sigma_i\sigma_j\ge-|J_{ij}|$
(equality iff $\sigma_i=\sigma_j$), so summing gives the bound with equality exactly at
$\bm\sigma=\mathrm{all}{+}1$. Independents therefore introduce no competing interaction and no
ground-state disorder: on the connected component containing any partisan voter, they are forced to
$\sigma_i=+1$ at zero energy cost via their ferromagnetic bonds, exactly like every other spin.
(A connected sub-graph consisting \emph{only} of zero-field independents, disconnected from every
partisan voter, would retain a global-flip degeneracy — not the case here, since the network is built
as a single connected component.) Their realised opinion $s_i=c_i\sigma_i=c_i$ is effectively random
only because their own camp label $c_i$ was assigned at random, not because the ground-state search
itself is disordered. This directly explains Table~\ref{tab:balance}'s
bond satisfaction at $\nu=0.05$ ($\approx 1-\nu$ at every density, within the field-dominated
corrections this subpopulation and non-zero $r_h$ introduce): a bond is satisfied in the ground
state whenever its realised sign matches the gauge-trivial value $c_ic_j$, which happens with
probability $1-\nu$ by construction (Eq.~\eqref{eq:balance-rule}) for a ground state close to
$s^*=c$. None of this is specific to the empirical calibration used here; it is a structural property
of any balanced signing on any graph. What it is \emph{not} a property of is the comparison this
paper actually makes: the independent baseline at $\alpha=0.50$ is a genuine $\pm J$ spin glass ---
i.i.d.\ random signs are not, in general, representable as $c_ic_j$ for any camp assignment, and
carry real, non-gauge-removable frustration whenever a cycle's sign-product is negative --- an odd
number of negative edges around that cycle, which can occur on a cycle of any length, not only odd
ones (Harary's balance theorem \cite{harary1953} is a statement about sign-products around cycles,
not cycle length). The question this paper
answers is therefore not whether $\nu=0$ is ordered (it must be), but how far that order, and its
statistical signature in the full sampled energy-gap distribution, extends as $\nu$ moves away from that known
limit toward the genuinely frustrated baseline.

\subsection{Sampling protocol and a schedule-convention identity}
\label{sec:schedule}

Writing the field-scaled (``scale$_h$'') and coupling-scaled (``scale$_J$'') conventions as
\begin{align}
H_h(\mathbf{s};r_h) &= -r_h\sum_i h_i s_i - \sum_{(i,j)}J_{ij}s_is_j, \label{eq:Hh}\\
H_J(\mathbf{s};r_h) &= -\sum_i h_i s_i - \frac{1}{r_h}\sum_{(i,j)}J_{ij}s_is_j, \label{eq:HJ}
\end{align}
both specialisations of Eq.~\eqref{eq:hamiltonian}'s sign convention at $r_h=1$ up to the stated
rescaling,
\begin{equation}
\operatorname*{arg\,min}_{\mathbf{s}} H_h(\mathbf{s};r_h) \;=\; \operatorname*{arg\,min}_{\mathbf{s}} H_J(\mathbf{s};r_h)
\label{eq:gs-invariance}
\end{equation}
for any $r_h>0$, since $H_h = r_h\,H_J$ and multiplying by a positive constant never changes which
configuration minimises it. This proves the ground-state \emph{configuration} is invariant between
conventions, and hence so are any statistics computed purely from that configuration --- frozen
fraction, bond satisfaction, and field share of ground-state energy (a ratio, where the common factor
$r_h$ cancels). \textbf{The minimum energy \emph{value} is not invariant}: $H_h(\mathbf{s}^*;r_h) =
r_h\,H_J(\mathbf{s}^*;r_h)$ at the shared minimiser $\mathbf{s}^*$, so raw ground-state energy scales
with $r_h$ between conventions and is excluded from this invariance claim. We confirmed both halves
directly (`empirical\_calibration/r\_h\_r\_j\_symmetry\_check.py`, paired same-disorder-realisation
Wilcoxon signed-rank): frozen fraction, bond satisfaction, and field share show no significant
divergence at any tested grid point, while raw $E_0/N$ differs significantly at every tested
$r_h\neq1$ (e.g.\ $m=5$, $r_h=4$: $-4.75$ vs.\ $-1.19$, a factor of exactly $4=r_h$; $p=1.9\times10^{-6}$,
the Wilcoxon floor at $n=20$), exactly as Eq.~\eqref{eq:gs-invariance}'s own scaling predicts.

\begin{figure}[htbp]
\centering
\includegraphics[width=\columnwidth]{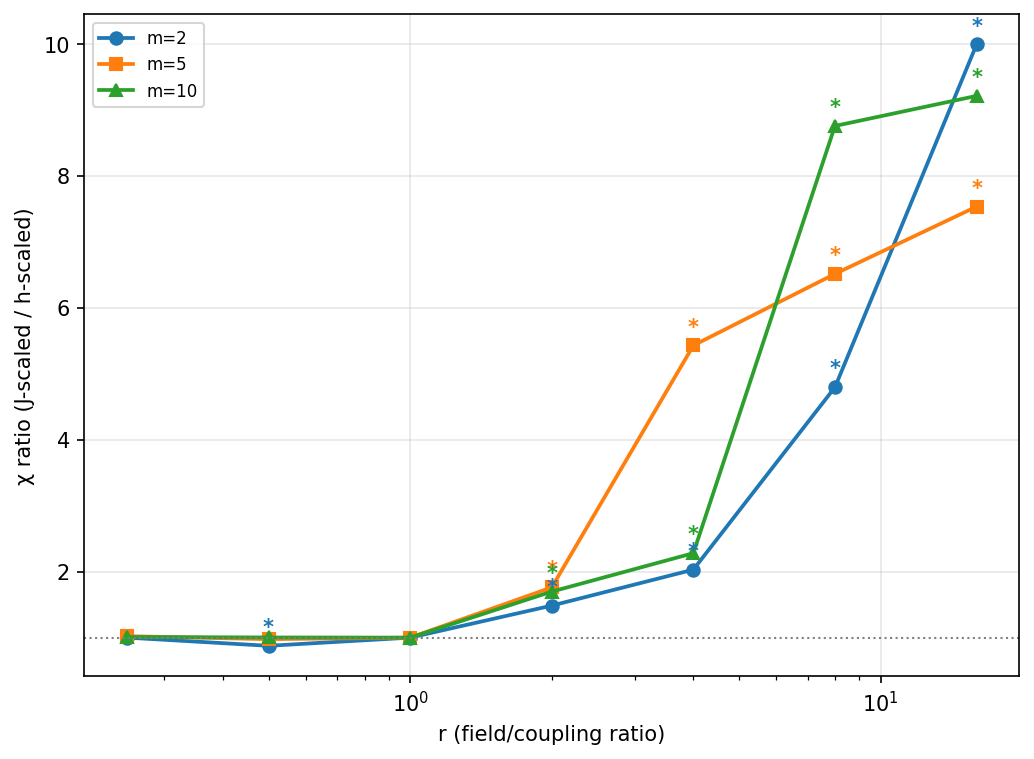}
\caption{Sampling-protocol invariance check. $\chi$ (susceptibility) ratio between the two schedule
constructions, by density ($n=20$ paired disorder realisations/cell, Wilcoxon signed-rank test,
$*=p<0.05$) --- diverges sharply from $r\approx2$, reaching $7.5$--$10\times$ by
$r=16$. Ground-state statistics are invariant to this choice by construction (Eq.~\eqref{eq:gs-invariance})
and are not plotted here, since a flat line at $1.0$ adds nothing beyond that statement. This
divergence is directly load-bearing for Sec.~\ref{sec:results}'s interpretation of the energy-gap
test's apparent weakening at high $r_h$ under the \emph{unadjusted} convention.}
\label{fig:schedule}
\end{figure}

We emphasise that simulated annealing is not equilibrium sampling at a defined physical temperature:
$\beta$ here is an annealing-schedule control parameter, swept over a fixed, finite range chosen for
search performance, not a thermodynamic inverse temperature, and $\chi=N\,\mathrm{Var}(m)$
(Sec.~\ref{sec:results}) is accordingly reported as a population-diversity proxy computed from the
resulting read population, not a susceptibility in the fluctuation-dissipation sense at a stated
equilibrium temperature. With that caveat stated plainly: statistics computed from the \emph{full}
sampled population rather than a single read --- $\chi$, and the Kolmogorov--Smirnov
sampled-energy-gap-distribution comparison used below --- require one further check, since
Eq.~\eqref{eq:gs-invariance} says nothing about the finite-$\beta$ distribution the annealer actually
samples along the way. At a given $\beta$, the proposal/acceptance step targets
$\mathbf{s}\propto\exp(-\beta H)$ as its stationary distribution --- matching $\beta H$ between two
constructions matches the transition probabilities at every step, given the same proposal, update
rule, initialisation, and schedule, but does not by itself establish that the finite-length anneal has
equilibrated to that distribution. Because the annealing schedule's $\beta$ range is fixed in absolute
units, scaling $h$ up at that
fixed schedule need not sample the same relative population diversity as scaling $J$ down at that
schedule, even though both reach the identical ground state. The two constructions agree closely for
$r_h \le 1$ but diverge substantially and non-saturatingly above it in our own tests of this
instance (Fig.~\ref{fig:schedule}). \textbf{They are, however, related by an
exact identity once the schedule itself is co-scaled} --- a one-line consequence of
$\exp(-\beta H)$'s linearity in $\beta H$, noted here because it resolves an apparent $r_h$-dependent
fade in Sec.~\ref{sec:results}'s energy-gap signal, not because the identity itself is
mathematically deep. Compensating scale$_h$'s schedule by
$\beta\to\beta_0/r_h$ (``scale$_h$ compensated''), relative to the nominal, unmodified schedule
$\beta_0$ used for scale$_J$,
\begin{align}
\frac{\beta_0}{r_h}\,H_h(\mathbf{s};r_h) &= -\beta_0\sum_i h_i s_i - \frac{\beta_0}{r_h}\sum_{(i,j)}J_{ij}s_is_j \notag\\
&= \beta_0\,H_J(\mathbf{s};r_h),
\label{eq:schedule-identity}
\end{align}
so scale$_h$ compensated and scale$_J$ assign the \emph{identical} Boltzmann weight
$\exp(-\beta_0 H_J)$ to every configuration $\mathbf{s}$, at every step of the anneal, not only at
the ground state --- confirmed numerically as bit-identical aggregate statistics to the last
reported decimal across the full tested grid. All full-population statistics reported here at
$r_h \ge 2$ use this schedule-adjusted protocol.

\section{Results}
\label{sec:results}

\begin{figure*}[htbp]
\centering
\includegraphics[width=\textwidth]{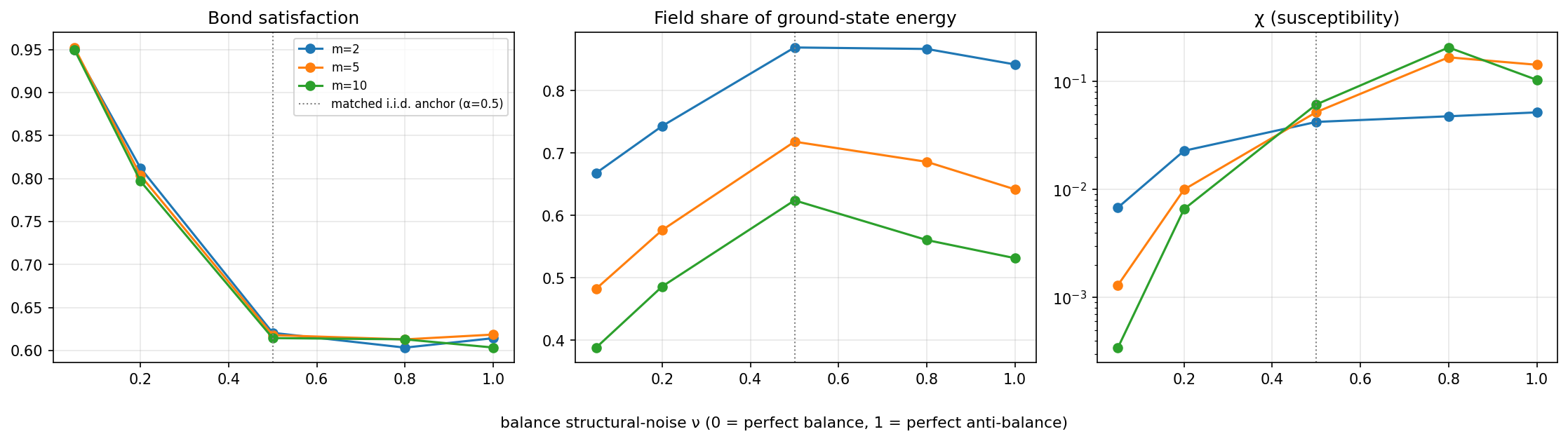}
\caption{Balance-structured vs.\ independent sign placement, $r_h=1$. Bond satisfaction, field
share, and $\chi$ vs.\ balance structural-noise ($0$ = perfect balance, $1$ = perfect anti-balance),
by density; dotted line marks the matched independent-sign anchor.}
\label{fig:balance}
\end{figure*}

Three statistics, each read from a run's lowest-energy sample $\mathbf{s}^*$, anchor this section.
Field share $F$ and bond satisfaction $S$ are
\begin{align}
F &= \frac{\big|\sum_i h_i s_i^*\big|}{\big|\sum_i h_i s_i^*\big| + \big|\sum_{(i,j)} J_{ij}s_i^*s_j^*\big|}, \label{eq:observables-f}\\
S &= \frac{1}{|E|}\sum_{(i,j)\in E} \mathbb{1}\!\left[J_{ij}s_i^*s_j^*>0\right], \label{eq:observables-s}
\end{align}
the field's share of total ground-state energy and the fraction of edges whose bond is satisfied
(aligned with its own sign) in the ground state, respectively; the field-dominance crossover is the
$r_h$ at which $F$ passes $50\%$. Susceptibility $\chi = N\,\mathrm{Var}(m)$, with
$m=\frac{1}{N}\sum_i s_i$ the per-read magnetisation and the variance taken over the full read
population rather than a single sample, measures ensemble spread. Moving from independent sign
placement ($\alpha=0.50$, matched marginal rate) to near-perfect Heider balance
($\nu=0.05$) at $r_h=1$ changes all three substantially (Table~I;
Fig.~\ref{fig:balance}): bond satisfaction rises by roughly 33 percentage points at every density
(balanced triads are jointly satisfiable; independent signs at $50\%$ negative are not); the
field-dominance crossover shifts $2$--$2.7\times$ ($0.75\to2.00$ at $m=5$, $1.00\to2.00$ at $m=10$);
and susceptibility --- reduced variation in sampled magnetisation across the read population, not a
thermodynamic susceptibility at a defined temperature (Sec.~\ref{sec:schedule}) --- collapses by an
order of magnitude or more. At the anti-balanced extreme ($\nu\to1.0$), bond satisfaction
falls \emph{below} the independent-sign baseline and susceptibility rises well above it --- a
frustrated population (anti-balance is a well-defined structural extreme, not shown here to be the
graph's globally most-frustrated signing) is both less bond-satisfied and more variable across samples than
unstructured disorder, not simply a mirror image of the balanced case, consistent with dynamical
structural-balance results showing unbalanced configurations are inherently unstable rather than
persisting as balance's opposite \cite{marvel2011}.

\begin{table*}[htbp]
\caption{Balance-structured vs.\ independent sign placement at $r_h=1$.}
\label{tab:balance}
\begin{ruledtabular}
\begin{tabular}{cccc}
$m$ & bond satisfaction & field share & $\chi$ \\
\hline
2  & $62.1\% \to 95.0\%$ & $86.9\% \to 66.8\%$ & $0.042 \to 0.007$ \\
5  & $61.8\% \to 95.3\%$ & $71.8\% \to 48.3\%$ & $0.052 \to 0.001$ \\
10 & $61.5\% \to 95.0\%$ & $62.4\% \to 38.8\%$ & $0.061 \to 0.0003$ \\
\end{tabular}
\end{ruledtabular}
\end{table*}

To test whether this landscape difference is statistically detectable in a genuinely different
sampled energy-gap distribution, rather than merely a mean-shift in ground-state statistics, we compare
balance-structured and independent sign placement using the confound-controlled design of Sec.~\ref{sec:sign}:
per realisation, one graph, one field assignment, and one shared set of $|J|$ edge magnitudes, with
balance-structured signs (near-perfect Heider balance) as the ``real'' condition against two
independent draws at the same matched marginal rate as the ``null'' pair. Energy gap,
$\Delta E = E - E_0$ with $E_0$ that run's own minimum energy, removes a raw-energy location-shift
confound. At the
empirical anchor ($r_h=1$, $N=90$, $m\in\{2,5,10\}$, 30 realisations per density), real-vs-null
Kolmogorov--Smirnov \cite{kolmogorov1933,smirnov1948} statistics run systematically higher than
null-vs-null (mean $D_{KS}=0.515$ vs.\ $0.356$; Mann--Whitney $p=7.3\times10^{-8}$), holding
individually at every density ($p=0.0012$, $0.0024$, $0.00069$ for $m=2,5,10$ --- surviving
Holm--Bonferroni correction for multiple comparisons both within this family and combined with the
$r_h$-sweep family below). This is confirmed at full production sampling resolution (5000 reads/2000
sweeps rather than 1000/1500): mean $D_{KS}$ essentially unchanged, Mann--Whitney
$p=1.46\times10^{-7}$. The reject-side separation is modest in absolute terms --- null-vs-null draws
from the \emph{same} i.i.d.\ construction already give mean $D_{KS}=0.356$, a real noise floor set by
sampler and finite-$N$ variability, against which real-vs-null's $0.515$ is a $\sim\!45\%$ relative
increase --- and the $p$-value should be read against that floor, not in isolation. We also note that
the real-vs-null and null-vs-null arms share one of their two draws per realisation (both are
compared against the same null$_1$ instance), so successive realisations are independent of each
other but the two arms are not fully independent \emph{within} a realisation; the Mann--Whitney test
above treats the pooled 30+30 values as unpaired, which is not a pairing-aware test (we do not assert
this is conservative with respect to power in general, only report the paired re-analysis directly
below rather than relying on that assumption). Re-analysing the same 90 cells with a per-realisation paired Wilcoxon
signed-rank test on each cell's own $D_{KS}^{\mathrm{real}}-D_{KS}^{\mathrm{null}}$ difference
confirms the result at comparable significance: $W=724.5$, $p=1.02\times10^{-7}$ (firmed-up settings)
and $W=762.0$, $p=2.31\times10^{-7}$ (production settings), individually significant at every density
($p=0.0026$/$0.0037$, $0.0071$/$0.0106$, $5.6\times10^{-5}$/$9.9\times10^{-5}$ for $m=2,5,10$,
firmed-up/production), positive in $81.1\%$/$78.9\%$ of cells. The more defensible test confirms,
rather than weakens, the finding.

\subsection{Isolating field--camp alignment from triadic structure}
\label{sec:isolate}

The balance-structured condition above assigns camp(i) from each voter's own real partisan-lean tier
(Sec.~\ref{sec:sign}), so that $\mathrm{sign}(h_i)=c_i$ exactly for every partisan voter
(Sec.~\ref{sec:gauge}). Its bonds are therefore correlated with field alignment in a way the
independent baseline's i.i.d.\ bonds are not, so the result above reflects this complete signing
construction, not necessarily triadic balance structure considered alone. We isolate them: take the
real camp assignment for a realisation and \emph{randomly permute it across voters}, reusing the same
realised edge-flip noise $\xi_{ij}$ (Eq.~\eqref{eq:balance-rule}) for both the aligned and permuted
camp assignments. Writing $\mathrm{sign}(J_{ij})=\xi_{ij}c_ic_j$, the product of $c_ic_j$ around any
closed cycle equals exactly $1$ for \emph{any} camp assignment (each $c_i$ appears exactly twice), so
every cycle's sign product --- and therefore triadic balance --- is identical between the aligned and
permuted conditions \emph{by construction}, independent of which specific camp assignment is used;
sharing $\xi_{ij}$ makes this an exact identity rather than an approximate match confirmed only by
measurement. The permutation reduces the correlation between camp and each voter's own field sign to
approximately zero (measured: $1.0$ exactly for the aligned condition, by construction, vs.\ $<0.10$
for the permuted one), while the realised marginal negative-edge rate is closely matched between the
two ($49.8\%$ aligned vs.\ $48.7\%$ permuted, averaged over tested cells).

Re-running the identical confound-controlled design with this field-independent condition in place of
the original, at the same empirical anchor ($r_h=1$, $N=90$, $m\in\{2,5,10\}$, 30 realisations/density,
sharing the same graph/field/edge-magnitude/noise draws as the original three arms), gives four
energy-gap comparisons per realisation: null-vs-null (negative control, A), field-aligned-vs-null
(B, the original result above), field-independent-permuted-vs-null (C, the isolating test), and
aligned-vs-permuted (D, direct).

At every density, no significant difference is detected between the permuted condition and the
negative control (C vs.\ A: Mann--Whitney $p=1.0,0.22,0.60$ for $m=2,5,10$, or $p=0.74$ pooled across
densities; pooled paired mean
difference $0.003$, 95\% bootstrap CI $[-0.041,0.046]$ --- disorder realisation, not pooled
energy-gap pair, is the statistical unit for the paired estimate and CI), while the field-aligned
condition remains sharply different from the permuted condition both unpaired (B vs.\ C:
Mann--Whitney $p=0.0032$, $5.9\times10^{-4}$, $2.3\times10^{-3}$) and, using the same-realisation
pairing both conditions share by construction, paired (D, aligned-vs-permuted: Wilcoxon signed-rank
$p=1.7\times10^{-16}$ pooled across densities) --- \textbf{D is the same B-vs-C comparison tested
with pairing, not a comparison against the negative control.}
Mean $D_{KS}$ by arm, pooled across densities: A$=0.356$, B$=0.521$, C$=0.359$. This does not establish
that the permuted condition is equivalent to the independent baseline --- only that no difference was
detected at this sample size and anchor --- but it does show that triadic structure held fixed at the
same level as the aligned condition, with its intrinsic bond frustration provably unchanged by the
permutation, is not sufficient on its own to reproduce the signal detected in (B).

\begin{table}[htbp]
\caption{Isolating field--camp alignment from triadic structure: mean $D_{KS}$ by arm and density, and
Mann--Whitney significance (A: null-vs-null; B: field-aligned; C: field-independent permuted).}
\label{tab:isolation}
\begin{ruledtabular}
\begin{tabular}{ccccccc}
$m$ & A & B & C & $p$(A vs.\ C) & $p$(B vs.\ C) \\
\hline
2  & 0.293 & 0.455 & 0.312 & 1.00 & $3.2\times10^{-3}$ \\
5  & 0.372 & 0.534 & 0.335 & 0.22 & $5.9\times10^{-4}$ \\
10 & 0.404 & 0.575 & 0.431 & 0.60 & $2.3\times10^{-3}$ \\
\end{tabular}
\end{ruledtabular}
\end{table}

This is not a cosmetic relabelling. A direct check of the single lowest-energy sample at the same
cells (not the full energy-gap distribution) shows substantially lower bond satisfaction in the
permuted condition's lowest observed configurations ($0.95\to0.64$ at $m=2$; $0.95\to0.78$ at $m=5$;
$0.95\to0.91$ at $m=10$), because field and bonds compete for the lowest-energy configuration once
camp no longer tracks field sign, even though the bond network's cycle structure --- and therefore its
intrinsic frustration in the standard graph-theoretic sense --- is, by the argument above, exactly
unchanged by the permutation. The drop in observed bond satisfaction is a statement about which
configuration the field-bond competition favours, not about the bond network's own frustration.
Parallel tempering (Sec.~\ref{sec:pt-extension}) reproduces the same pattern, subject to its own
reported convergence diagnostics --- no significant
difference detected between the permuted condition and the negative control, a sharp difference
between the aligned and permuted conditions --- so this result is
not an artefact of the finite annealing schedule used elsewhere in this paper.

\begin{figure*}[htbp]
\centering
\includegraphics[width=\textwidth]{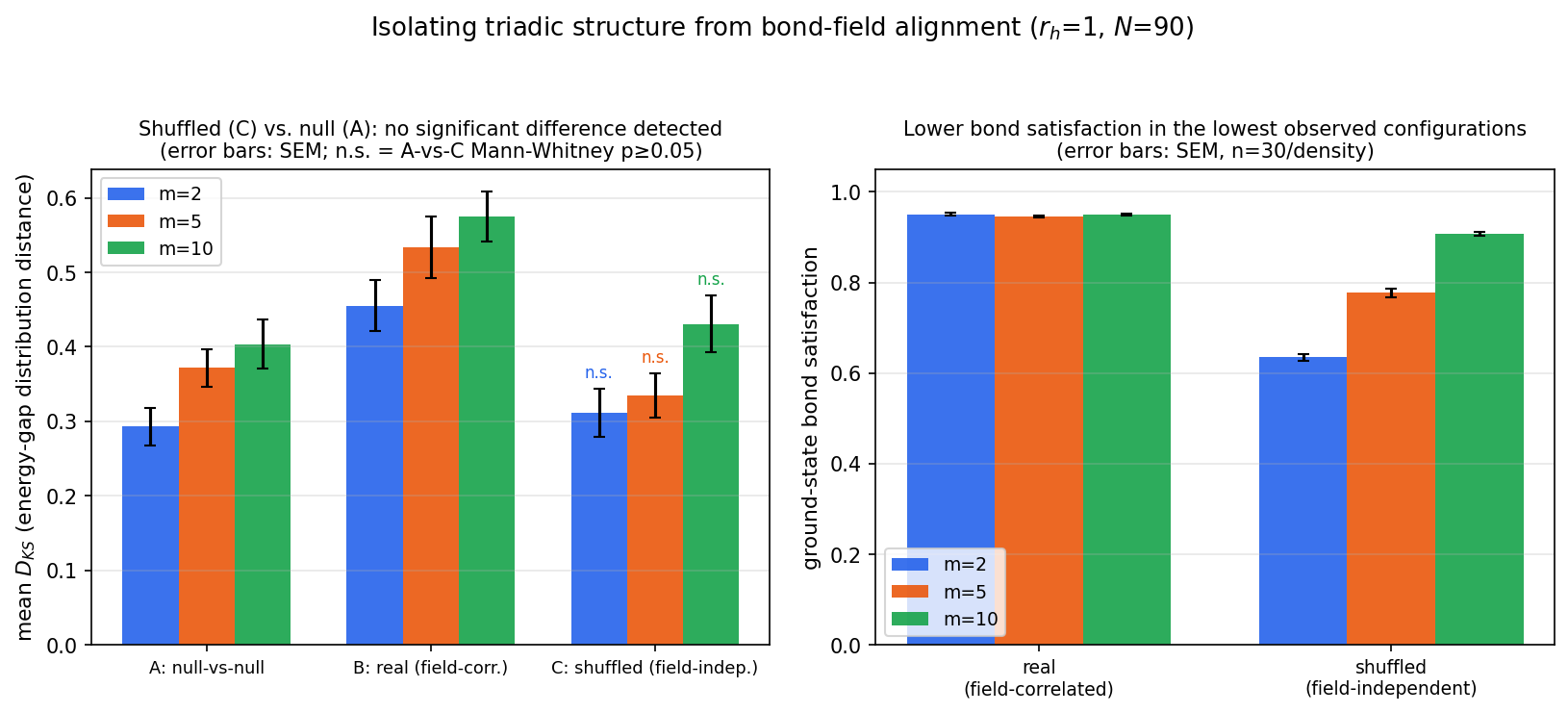}
\caption{Isolating field--camp alignment from triadic structure. Left: mean $D_{KS}$ by arm and
density (error bars: SEM over 30 realisations; n.s.\ marks the A-vs-C comparisons from Table
\ref{tab:isolation} where no significant difference from the negative control was detected). Right:
bond satisfaction in the single lowest observed configuration, aligned vs.\ permuted, by density.}
\label{fig:isolation}
\end{figure*}

As anticipated in the Introduction, results (2)/(3) above --- framed around structural balance
broadly --- in fact depend specifically on field--camp alignment, not triadic structure considered
on its own: triadic structure held fixed, on its own, does not reproduce a detectable signal at this
sample size, while field--camp \emph{alignment} does. The claim this paper supports is
that field--camp alignment --- which coincides with near-perfect triadic balance in the calibration
used throughout, since camp is drawn from real partisan lean --- is responsible for the detected
difference in sampled energy-gap distributions, not triadic balance considered on its own.

\subsection{Confirmation under parallel-tempering sampling}
\label{sec:pt-extension}

All results above are simulated annealing under a fixed, finite schedule (Sec.~\ref{sec:schedule}),
not equilibrium sampling at a defined temperature: $\beta$ is a schedule-control parameter, and
sampled magnetisation variance is correspondingly a population-diversity proxy, not a fluctuation-
dissipation susceptibility. We built a parallel-tempering (replica-exchange Monte Carlo) sampler ---
validated against exact enumeration on small systems --- and applied it directly to
both questions this raises, rather than leaving them disclosed-only. \textbf{Settings:} 12 replicas
on a geometric $\beta$ ladder from $0.1$
(hottest) to $10.0$ (coldest/target replica); 3000 burn-in sweeps followed by 6000 measurement
sweeps per run; magnetisation and energy sampled every 4th sweep during the measurement phase
(1500 samples/replica); convergence diagnosed by per-adjacent-pair swap-acceptance rate and a
first-half/second-half energy block check on the coldest replica. \textbf{Run-length stability
check:} a representative subset (8 disorder realisations at $m=10$, 8 at $m=2$, $r_h=1$, same seeds)
was additionally rerun at $2\times$ and $4\times$ this length. Comparing each realisation's own value
at $1\times$ against $4\times$ (paired Wilcoxon signed-rank, $n=16$): no significant difference for
$D_{KS}$(B, aligned-vs-null; $p=0.21$), $D_{KS}$(C, permuted-vs-null; $p=0.86$), $D_{KS}$(D,
aligned-vs-permuted, the headline direct comparison; $p=0.10$), or $\chi$ ($p=0.24$). The pooled mean
of $D_{KS}$(D) does shift from $0.429$ at $1\times$ to $0.521$ at $4\times$ --- a real point-estimate
change we report rather than hide --- but it is not statistically distinguishable from no change at
this sample size; $D_{KS}$(C) and $\chi$ are essentially unchanged in raw magnitude ($0.327\to0.343$
and $0.00417\to0.00404$ respectively). We read this as no significant instability detected under a
$4\times$ increase in PT length, not as proof the published settings have converged to the true
equilibrium distribution --- a non-significant result at $n=16$ cannot establish that, only that this
specific check did not detect drift.

At the empirical anchor ($N=90$, $m=10$, $r_h=1$, 30 realisations), equilibrium-sampled
$\chi=N\,\mathrm{Var}(m)$ reproduces the sampled-magnetisation-variance result above: the field-aligned
condition's $\chi$ (mean $0.00081$) is roughly $80\times$ smaller than the independent baseline's
(mean $0.0628$; Mann--Whitney $p=6.3\times10^{-11}$, paired Wilcoxon $p=1.9\times10^{-9}$), matching
the order of magnitude and direction of the annealing-based result at the same cell.

Extending equilibrium sampling to the isolation comparison itself (same four arms as
Sec.~\ref{sec:isolate}, same scale, $N$=90, $m\in\{2,5,10\}$, 30 realisations/density), the same
pattern holds: no significant difference is detected between the permuted condition and the negative
control at any density (A-vs-C: $p=0.65$, $0.16$, $0.67$ for $m=2,5,10$), while the field-aligned
condition remains sharply different from the permuted one (B-vs-C: $p=0.0025$, $2.8\times10^{-5}$,
$2.1\times10^{-4}$) and from the negative control by paired comparison ($p=1.7\times10^{-16}$ pooled).
The minimum per-adjacent-pair swap-acceptance rate across all runs was $3.8\%$, below this
implementation's own healthy-acceptance target in places (plausibly reflecting a sharper ordering
transition near the gauge-equivalent-ferromagnet limit); links adjacent to the coldest/target replica
remained healthy and block checks were stable in spot checks, so the target replica's own measurement
is judged reliable despite the weaker link elsewhere in the ladder. This result is not an artefact of
the finite annealing schedule used elsewhere in this paper.

\begin{figure*}[htbp]
\centering
\includegraphics[width=\textwidth,trim=0 0 0 29,clip]{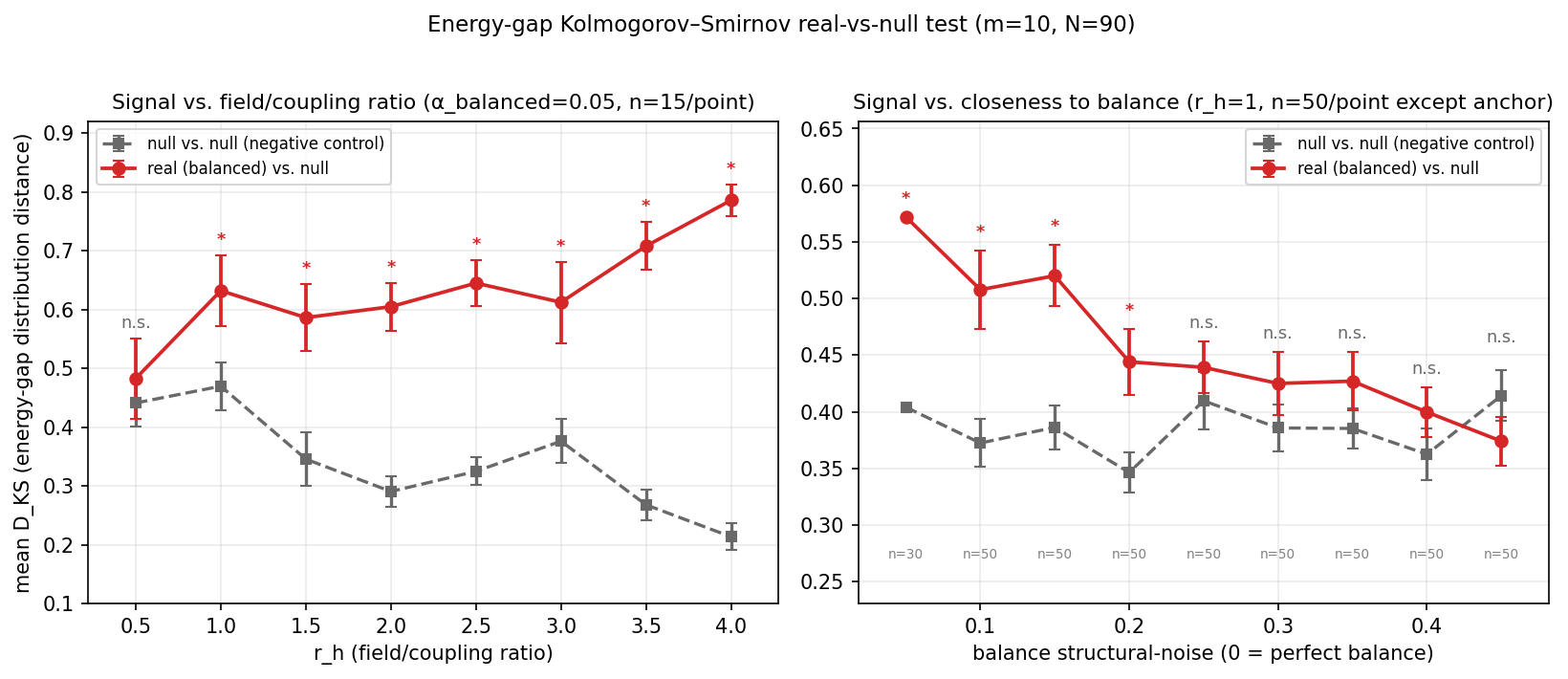}
\caption{Energy-gap Kolmogorov--Smirnov real-vs-null test ($m=10$, $N=90$). Left: significance
vs.\ $r_h$ under the schedule-adjusted sampling protocol. Right: significance vs.\ closeness to
balance at $r_h=1$; per-point sample size shown below each pair ($n=30$ at the reused anchor,
$n=50$ elsewhere), and the anchor point is plotted without an error bar since it reuses an earlier
pass that did not retain per-trial values. $* = p<0.05$.}
\label{fig:ks}
\end{figure*}

Sweeping $r_h$ under the schedule-adjusted sampling protocol of Sec.~\ref{sec:schedule} at 8 points,
$n=15$ trials/point (Fig.~\ref{fig:ks}, left; at $r_h=1.0$ the schedule-adjusted and unadjusted
constructions are mathematically identical by construction, so that point also serves as an internal
consistency check against the $n=30$ anchor above, $p=0.020$ vs.\ $p=0.00069$ --- both reject the
null; the two differ in $n$ and realisation draws, not in conclusion) finds no
significant difference at $r_h=0.5$ ($p=0.53$) but a significant difference at every point from
$r_h=1.0$ through $4.0$ ($p=0.020$, $0.0037$, $5.7\times10^{-5}$, $4.8\times10^{-5}$, $0.012$,
$6.1\times10^{-6}$, $3.4\times10^{-6}$) --- the real-vs-null gap widens overall as the
field/coupling ratio grows, though not perfectly monotonically: both the mean $D_{KS}$ and the
significance level dip slightly around $r_h=3.0$ before strengthening again through $r_h=4.0$, a
real feature the original 4-point grid was too coarse to resolve. Within this 8-point family
alone, Holm--Bonferroni correction leaves all 7 significant points standing. The same
schedule-relative-temperature effect responsible for the Sec.~\ref{sec:schedule} divergence at
$r_h\ge2$ also appears at $r_h=0.5$, running in the opposite direction: scaling $h$ down (rather
than up) makes the fixed schedule proportionally hotter, so the uncompensated construction explores
\emph{more} there and gives a stronger, significant signal ($p=0.0225$) than the schedule-adjusted
value reported above ($p=0.53$) --- one continuous, already-diagnosed mechanism (monotonic in
$r_h$ across the full tested range) rather than a separate issue confined to high $r_h$. We report
the schedule-adjusted value throughout, as our stated protocol requires. Sweeping
closeness-to-balance at fixed $r_h=1$ (Fig.~\ref{fig:ks}, right), $n=50$ trials/point, finds the
effect requires fairly close to genuine Heider balance: significant through $\nu=0.20$
($p=0.0038$, $0.00012$, $0.0069$) but not from $\nu=0.25$ onward
($p=0.24$--$0.37$). Within this 8-point family alone, Holm--Bonferroni correction leaves all 3
significant points standing. Pooling all three families tested in this paper (per-density,
$r_h$-sweep, closeness-to-balance --- 19 tests total) into a single, maximally conservative
Holm--Bonferroni frame, the step-down procedure's first failure occurs earlier, at
closeness-to-balance$=0.20$ (rank 11 of 19): every test ranked after it is treated as failed
regardless of its own $p$-value, including two points that survive comfortably within their own
family ($r_h=1.0$, $p=0.020$; $r_h=3.0$, $p=0.012$). 10 of 19 tests survive this maximally
conservative framing; we report both the within-family and the fully-pooled framing rather than
selecting whichever looks stronger.

\textbf{Calibrating $\nu$ against one measured real-network balance statistic.} Leskovec,
Huttenlocher \& Kleinberg \cite{leskovec2010} report the balanced-triad fraction directly for three
real signed networks: Epinions $0.941$, Slashdot $0.912$, Wikipedia $0.909$ --- all \emph{more}
balanced than this paper's own $\nu=0.05$ condition ($0.8645$). Inverting
$(1-\nu)^3+3\nu^2(1-\nu)$ gives the $\nu$ that reproduces \emph{this one statistic} exactly:
$\nu=0.0205$ (Epinions), $0.0312$ (Slashdot), $0.0324$ (Wikipedia). \textbf{This gives a
triangle-statistic-matched noise level, not a full network calibration}: it does not reproduce these
networks' marginal negative-edge rates, topology, longer-cycle structure, or field correlations, none
of which this paper's construction attempts to match. Re-running the original
confound-controlled comparison at each of these three values, same scale as
the anchor ($N=90$, $m\in\{2,5,10\}$, 30 realisations/density): mean $D_{KS}$(real$-$null), the
effect-size difference the significance tests below are computed from, is $0.174$ (Epinions), $0.165$
(Slashdot), $0.163$ (Wikipedia) --- each modestly larger than the original $\nu=0.05$ anchor's $0.159$
(a $2$--$9\%$ relative increase), alongside Mann--Whitney $p=5.7\times10^{-9}$ (Epinions),
$2.2\times10^{-8}$ (Slashdot), $2.7\times10^{-8}$ (Wikipedia) vs.\ the original's $7.3\times10^{-8}$,
individually significant at every density for all three networks
($p=0.0001$--$0.0017$, $0.0003$--$0.0024$, $0.0004$--$0.0022$ respectively). The much larger drop in
$p$-value than in effect size is expected, not itself evidence of a dramatically stronger effect:
lower $\nu$ means a more consistent, less noisy balanced structure, which reduces $p$ substantially
even when the underlying effect size grows only modestly. We report both effect size and significance
rather than the significance level alone. The arbitrary endpoint
used throughout this paper is, on this one statistic, a conservative choice relative to the three
networks checked.

The isolation comparisons of Sec.~\ref{sec:isolate} (B-vs-C, 3 densities) and the real-network-$\nu$
comparisons above (3 networks $\times$ 3 densities) are treated as two further, separate families
rather than pooled into the 19-test frame above: each tests a qualitatively different hypothesis ---
whether structure alone, alignment held fixed, differs from alignment present; and whether the
original anchor's conclusion is robust to using a real-network-calibrated $\nu$ instead of the
arbitrary $\nu=0.05$ --- not an additional sweep point along the $r_h$ or closeness-to-balance axes
already pooled. Holm--Bonferroni within each: all 3 B-vs-C tests survive ($p=0.0032$, $0.00059$,
$0.0023$); all 9 real-network-$\nu$ tests survive ($p=0.0001$--$0.0024$). The non-significant A-vs-C
comparisons (Sec.~\ref{sec:isolate}) are reported as point estimates with a confidence interval, not
as significance claims, so Holm--Bonferroni correction --- which guards against false positives, not
false negatives --- does not apply to them.

\section{Discussion}

The $\nu=0$ endpoint's near-total ground-state order is not, on its own, new physics:
Sec.~\ref{sec:gauge}'s gauge mapping, and the fact that $h_i$'s sign already matches camp membership
for every partisan voter in this calibration, together mean that limit is close to built into the
construction. Sec.~\ref{sec:isolate}'s isolating experiment shows that what matters for the sampled
energy-gap distribution, away from that trivial limit, is specifically the correlation between
camp and field sign --- not triadic balance structure considered on its own, which produces no
detectable signal when held fixed at the same level with that correlation removed.

What does this establish beyond what a ferromagnet with aligned versus competing transformed fields
already implies? Writing $\mathrm{sign}(J_{ij})=\xi_{ij}c_ic_j$ (shared
edge-flip noise $\xi_{ij}$, Sec.~\ref{sec:isolate}), the \emph{same} gauge transform
$\sigma_i=c_is_i$ applies at \emph{any} $\nu$, not only $\nu=0$: it sends the coupling term to
$-\sum\xi_{ij}|J_{ij}|\sigma_i\sigma_j$ for every $\nu$, since $c_i^2c_j^2=1$ regardless of how much
structural noise is present. Applying the \emph{same} construction to the permuted condition with
its own labels $c_i'$ (a permutation of the real camp) gives the \emph{identical} transformed
coupling pattern $-\sum\xi_{ij}|J_{ij}|\sigma_i\sigma_j$ --- confirmed directly, both algebraically
and numerically --- because the permuted condition reuses the same realised $\xi_{ij}$. The aligned
and permuted conditions' own gauge transforms therefore differ only in the transformed \emph{field}:
$c_ih_i=|h_i|$ (uniform sign) for the aligned condition, $c_i'h_i$ (a sign effectively random with
respect to that voter's own field, since $c_i'$ is an unrelated permutation) for the permuted one.
The aligned-vs-permuted comparison this paper reports is therefore, exactly, a comparison between
``the same realised bond disorder under a uniform field'' and ``the same realised bond disorder under
a random-sign field'' --- a known qualitative picture (uniform-field order vs.\ random-field
competition) that holds at every $\nu$ tested, not only at $\nu=0$. \textbf{The mapping alone
therefore already predicts the qualitative direction of Sec.~\ref{sec:isolate}'s result} (aligned
orders more than permuted) \textbf{before any simulation is run.} What the mapping does \emph{not}
predict, and what Sec.~\ref{sec:isolate} and
Sec.~\ref{sec:pt-extension} establish empirically rather than algebraically, is: (i) the effect's
\emph{size} in the specific statistic a Kolmogorov--Smirnov test on sampled energy gaps is
sensitive to, (ii) whether that size is large enough to separate cleanly from sampling noise at
$N=90$ --- a question about this finite system's noise floor, not about the gauge-equivalent
disorder, (iii) whether the effect survives parallel-tempering sampling rather than only the
simulated-annealing schedule used elsewhere, and (iv) whether triadic structure contributes anything
of its own once alignment is held fixed (no detectable contribution at $N=90$; at $N=180$--$300$, at
most a small one relative to alignment). The paper's contribution
is therefore not the qualitative discovery that field--camp alignment produces greater order ---
the gauge mapping already establishes that --- but the \emph{quantitative, confound-controlled
measurement} of how strongly alignment modifies the finite-system sampled energy-gap distribution
under controlled bond disorder, and the demonstration that this effect survives parallel-tempering
sampling and remains dominant over triadic balance alone at every tested system size
(Sec.~\ref{sec:results}).

The schedule-convention point of Sec.~\ref{sec:schedule} is included because it resolves what would
otherwise look like a puzzling, $r_h$-dependent fade in the real-vs-null signal as a measurement
artefact of one specific, common annealing-schedule convention rather than a genuine physical
weakening --- a caution relevant to any study of a field/coupling-ratio-swept Ising model under
simulated or quantum annealing --- but we do not present the underlying rescaling itself as a
theoretical result in its own right.

\textbf{Scope.} This paper isolates one question --- does sign-placement structure matter at fixed
marginal rate, away from the gauge-trivial balanced limit --- using an empirically-sourced field and
coupling magnitude as a fixed instance of the broader model, not as a claim about where a real
electorate sits on the swept $r_h$ axis. The independent baseline's $\alpha=0.50$ is also the i.i.d.\
construction's point of maximum mean-field frustration (the effective ferromagnetic coupling
$\langle k\rangle\langle|J|\rangle(1-2\alpha)$ vanishes exactly there), so the comparison reported
here is built at the sharpest available contrast between the two constructions, not at the marginal
rates --- roughly $15$--$23\%$ negative --- actually measured in Epinions, Slashdot, and Wikipedia
\cite{leskovec2010}. Because the balance-structured construction's own marginal negative rate stays
pinned near $50\%$ for every structural-noise value (camp membership is close to $50/50$ and
uncorrelated with graph structure, so it cannot be tuned down to $20\%$ by this construction), lowering
only the independent baseline's marginal rate to $0.20$ would compare a $\sim\!50\%$-negative balanced
condition against a $20\%$-negative independent one --- an \emph{unmatched} comparison that changes
marginal rate and structure simultaneously, not a matched-rate test at a realistic rate. (Such an
unmatched comparison does find a significant difference, Mann--Whitney $p=3.45\times10^{-12}$, a
genuine if differently-confounded empirical fact, but it does not establish matched-rate robustness at
a realistic marginal rate.) A genuinely matched-rate test at $\sim\!20\%$ would require a balanced
condition independently constructed to land near that rate, which the current construction cannot
produce; the real-network-calibrated $\nu$ comparison reported in Sec.~\ref{sec:results} addresses
the same underlying question --- does the effect hold at realistic, not arbitrary, structural-noise
levels --- in a way this construction \emph{can} support.

All results above use $N=90$. Re-running the original design at $N\in\{30,80,180\}$ (10
realisations per density, reusing the $N=90$/30-realisation result above rather than recomputing it)
finds no significant signal at $N=30$ (null and real-vs-null mean $D_{KS}$ essentially identical,
$0.274$ vs.\ $0.277$; Mann--Whitney $p=0.994$; mean difference $0.0036$, 95\% bootstrap CI
$[-0.066,0.075]$ --- a wide interval straddling zero, consistent with either a true null or a real
effect comparable to $N=80$'s; this is underpowered at $n=10$/density, not confident evidence of
absence), a signal that emerges by $N=80$ ($p=7.1\times10^{-4}$) and strengthens through $N=180$
($p=7.1\times10^{-9}$, individually significant at every density even at this reduced power), with
real-vs-null mean $D_{KS}$ rising monotonically ($0.277\to0.473\to0.515\to0.683$) as $N$ grows. The
headline $N=90$ result is therefore not a small-system artefact that would wash out at larger $N$ ---
if anything the effect strengthens with system size --- though it does not hold at every tested $N$,
and we report the $N=30$ null plainly rather than omitting it.

The isolation design's own central comparison (Sec.~\ref{sec:isolate}) --- whether triadic structure
alone (C) remains indistinguishable from noise (A) --- was checked at two further sizes using the
same confound-controlled four-arm design, at full power (30 realisations/density, matching the
$N=90$ anchor rather than the reduced $n=10$ above). The positive control held at both sizes: B
(field-aligned) remained sharply different from C at $N=180$ and $N=300$ alike (Mann--Whitney
$p=3.2\times10^{-23}$ and $p=2.1\times10^{-24}$ respectively), ruling out the alternative explanation
that the test simply lost power at larger $N$. Against that backdrop, neither formal test detects a
C-vs-A difference at either size (Mann--Whitney $p=0.12$ and $0.53$; paired Wilcoxon $p=0.13$ and
$0.44$, at $N=180$ and $N=300$ respectively), non-monotonically relative to the anchor's own pooled
$p=0.74$ (Sec.~\ref{sec:results}). The paired mean difference (C$-$A, same realisations and seeds
shared by both arms) is, however, consistently positive above the anchor: $0.045$ at $N=180$ (95\%
bootstrap CI $[0.001,0.089]$) and $0.037$ at $N=300$ (95\% CI $[-0.002,0.078]$), against $0.003$ at
$N=90$. We therefore do not claim the balance-only contribution is exactly zero at larger $N$: the
data are compatible with a small structural effect of up to $\approx0.09$ in $D_{KS}$, compared with
an alignment effect (B$-$A) of $0.431$ ($N=180$) and $0.518$ ($N=300$) at the same sizes ---
roughly an order of magnitude larger. Field--camp alignment remains the dominant contribution at
every tested size, consistent with the point estimates never approaching the field-alignment
effect's own scale even where the structural-only estimate is nominally positive; this is also why
the field-alignment effect above was itself invisible at $N=30$ and only emerged by $N=80$, while the
much smaller structural-only estimate here never reaches significance at any tested $N$. Realised mean
degree at $m=10$, measured independently with this check's own seeds (30 realisations/density,
same convention as Sec.~\ref{sec:network}), is $11.48$ ($N=180$) and $11.04$ ($N=300$), continuing
the downward trend already reported there ($12.37$ at $N=90$) as degree-cap pruning removes a larger
edge fraction at bigger $N$ --- a modest confound disclosed alongside this check rather than folded
silently into the result.

This model's broader phase
structure, energy landscape, and quantum-hardware behaviour away from the sign-placement question are
reported separately \cite{romero2026row2}.

The results above are subject to several limitations, beyond the statistical-independence caveat
already noted in Sec.~\ref{sec:results}. Camp membership for the balance-structured rule is assigned
from real partisan-lean tiers, but is close to $50/50$ and uncorrelated with graph structure by
construction; a real signed network's camp structure may correlate with degree or community structure
in ways not modelled here. The independent-vs-balanced comparison is matched in marginal rate at only
one point ($\alpha=0.50$ independent vs.\ $\nu=0.05$ near-perfect balance, both $\sim\!50\%$ negative);
the $\alpha=0.20$ independent re-run discussed in this section's Scope paragraph above is \emph{not}
a matched-rate comparison and should not be read as one. A genuinely matched-rate
comparison at a realistic ($\sim\!20\%$) marginal rate remains future work; calibrating $\nu$ against
measured real-network balance statistics is now done (Sec.~\ref{sec:results}) and strengthens, not
weakens, the finding. (Camp labels for the balance-structured rule
coincide with each partisan voter's own field sign by construction, Sec.~\ref{sec:sign};
Sec.~\ref{sec:isolate} isolates this correlation directly and shows that field--camp alignment,
rather than triadic structure alone, is the primary determinant of the reported effect.) The
finite-size sweep above covers
$N\in\{30,80,90,180\}$ only (10 realisations per density at the three new points, 30 at the $N=90$
anchor) for the original two-arm comparison; the isolation design's own C-vs-A comparison was
additionally checked up to $N=300$ (above), but $N>300$ remains untested throughout. Most results in this paper use simulated
annealing under a finite schedule, not equilibrium methods at a defined temperature or exact ground
states; Sec.~\ref{sec:schedule} addresses one specific schedule-convention artefact this introduces.
The exception is Sec.~\ref{sec:pt-extension}, which reruns the central $\chi$ comparison and the full
KS isolation test under parallel-tempering sampling at a defined temperature ---
a check of those specific results under a genuinely different sampler, not a substitute for one across every result
in this paper: the finite-size sweep, the $r_h$/closeness-to-balance sweeps, and the real-network-$\nu$
calibration above remain simulated-annealing-only. The edge-type frequency distribution underlying
$J$'s calibration (Sec.~\ref{sec:calibration}) is drawn from a fixed assumption rather than an independently measured
distribution.

\section*{Data and Code Availability}

The simulation code, raw result files, and figure-generation scripts supporting this paper are
available from the corresponding author upon request. The associated repository is not
yet publicly released while submission to a peer-reviewed journal is in preparation. Every reported
figure and table is reproducible directly from the cited data files using the corresponding analysis
script.


\begin{thebibliography}{99}

\bibitem{binder1986} K.~Binder and A.~P.~Young, Spin glasses: Experimental facts, theoretical concepts, and open questions, Rev. Mod. Phys. \textbf{58}, 801 (1986).
\bibitem{mezard1987} M.~M\'ezard, G.~Parisi, and M.~A.~Virasoro, \emph{Spin Glass Theory and Beyond} (World Scientific, 1987).
\bibitem{galam2004} S.~Galam, Contrarian deterministic effects on opinion dynamics: the hung elections scenario, Physica A \textbf{333}, 453 (2004).
\bibitem{castellano2009} C.~Castellano, S.~Fortunato, and V.~Loreto, Statistical physics of social dynamics, Rev. Mod. Phys. \textbf{81}, 591 (2009).
\bibitem{leskovec2010} J.~Leskovec, D.~Huttenlocher, and J.~Kleinberg, Signed networks in social media, in \emph{Proc. CHI 2010} (ACM, 2010), pp.~1361--1370.
\bibitem{heider1946} F.~Heider, Attitudes and cognitive organization, J. Psychol. \textbf{21}, 107 (1946).
\bibitem{cartwright1956} D.~Cartwright and F.~Harary, Structural balance: a generalization of Heider's theory, Psychol. Rev. \textbf{63}, 277 (1956).
\bibitem{harary1953} F.~Harary, On the notion of balance of a signed graph, Michigan Math. J. \textbf{2}, 143 (1953).
\bibitem{romero2026row2} N.~Romero, Social-Ising Model: Network Density and the Field--Coupling Ratio Jointly Determine Phase Structure and Landscape Geometry, in preparation (2026).
\bibitem{mattis1976} D.~C.~Mattis, Solvable spin systems with random interactions, Phys. Lett. A \textbf{56}, 421 (1976).
\bibitem{toulouse1977} G.~Toulouse, Theory of the frustration effect in spin glasses: I, Commun. Phys. \textbf{2}, 115 (1977).
\bibitem{facchetti2011} G.~Facchetti, G.~Iacono, and C.~Altafini, Computing global structural balance in large-scale signed social networks, Proc. Natl. Acad. Sci. \textbf{108}, 20953 (2011).
\bibitem{macy2024} M.~W.~Macy, B.~K.~Szymanski, and J.~A.~Ho{\l}yst, The Ising model celebrates a century of interdisciplinary contributions, npj Complexity \textbf{1}, 10 (2024).
\bibitem{korbel2023} J.~Korbel, S.~D.~Lindner, T.~M.~Pham, R.~Hanel, and S.~Thurner, Homophily-based social group formation in a spin glass self-assembly framework, Phys. Rev. Lett. \textbf{130}, 057401 (2023).
\bibitem{pham2022} T.~M.~Pham, J.~Korbel, R.~Hanel, and S.~Thurner, Empirical social triad statistics can be explained with dyadic homophylic interactions, Proc. Natl. Acad. Sci. \textbf{119}, e2121103119 (2022).
\bibitem{kirkpatrick1983} S.~Kirkpatrick, C.~D.~Gelatt, and M.~P.~Vecchi, Optimization by simulated annealing, Science \textbf{220}, 671 (1983).
\bibitem{bartels2000} L.~M.~Bartels, Partisanship and voting behavior, 1952--1996, Am. J. Polit. Sci. \textbf{44}, 35 (2000).
\bibitem{nickerson2008} D.~W.~Nickerson, Is voting contagious? Evidence from two field experiments, Am. Polit. Sci. Rev. \textbf{102}, 49 (2008).
\bibitem{long1997} J.~S.~Long, \emph{Regression Models for Categorical and Limited Dependent Variables} (Sage, 1997).
\bibitem{keith1992} B.~E.~Keith, D.~B.~Magleby, C.~J.~Nelson, E.~A.~Orr, M.~C.~Westlye, and R.~E.~Wolfinger, \emph{The Myth of the Independent Voter} (University of California Press, 1992).
\bibitem{anes2020} American National Election Studies, \emph{ANES 2020 Time Series Study Full Release} [dataset] (University of Michigan, 2021).
\bibitem{gerber2008} A.~S.~Gerber, D.~P.~Green, and C.~W.~Larimer, Social pressure and voter turnout: Evidence from a large-scale field experiment, Am. Polit. Sci. Rev. \textbf{102}, 33 (2008).
\bibitem{bond2012} R.~M.~Bond \emph{et al.}, A 61-million-person experiment in social influence and political mobilization, Nature \textbf{489}, 295 (2012).
\bibitem{granovetter1973} M.~S.~Granovetter, The strength of weak ties, Am. J. Sociol. \textbf{78}, 1360 (1973).
\bibitem{barabasi1999} A.-L.~Barab\'asi and R.~Albert, Emergence of scaling in random networks, Science \textbf{286}, 509 (1999).
\bibitem{marvel2011} S.~A.~Marvel, J.~Kleinberg, R.~D.~Kleinberg, and S.~H.~Strogatz, Continuous-time model of structural balance, Proc. Natl. Acad. Sci. \textbf{108}, 1771 (2011).
\bibitem{kolmogorov1933} A.~Kolmogorov, Sulla determinazione empirica di una legge di distribuzione, G. Ist. Ital. Attuari \textbf{4}, 83 (1933).
\bibitem{smirnov1948} N.~Smirnov, Table for estimating the goodness of fit of empirical distributions, Ann. Math. Stat. \textbf{19}, 279 (1948).

\end{thebibliography}
\end{document}